# From Vintage Mythology to Topological Physics: Unveiling a Universal Structural Attractor in Alcoholic Beverage Aging

Xinyue Jiang[1], Heng Yang[2], Zhiyin Jiu[3], Youxi Luo[3], Lin Chen[4], Yuqun Xie[2,*]

[1]School of Civil Engineering Architecture and Environment, Hubei University of Technology, Wuhan 430068, China

[2]School of Life and Health Sciences, Hubei University of Technology, Wuhan 430068, China

[3]School of Sciences, Hubei University of Technology, Wuhan 430068, China

[4]School of Art and Design, Hubei University of Technology, Wuhan 430068, China

## Summary

Alcoholic beverage properties are increasingly understood through ethanol-water structural states rather than empirical labels such as alcohol content and vintage. Yet whether chronological vintage similarly reflects an intrinsic structural state remains unclear. Here, we apply persistent homology to map the topological evolution of self-assembled molecular aggregates in strong-aroma Baijiu aged 1-10 years. The resulting fingerprints reveal a three-stage maturation pathway: rapid scaffold consolidation ($\beta_0$), population-level channel stabilization ($\beta_1$), and non-monotonic cavity reorganization ($\beta_2$). These coupled trajectories converge toward a mature topological state rather than passively tracking chronological age. We therefore propose a universal topological attractor, in which optimal aging is defined by a system's position in persistence space relative to a mature structural domain. This framework reframes beverage aging as navigation through structural state space, providing a physical basis for quality evaluation and accelerated maturation.

## Introduction

Since Einstein's annus mirabilis in 1905, modern physics has recast time from an absolute background into a coordinate that describes material states and their evolution. Consequently, any quality metric anchored solely in temporal duration constitutes an empirical convention rather than an intrinsic physical descriptor. Nevertheless, in the centuries-old craft of alcoholic beverage production, “vintage,” the mere accumulation of years, persistently serves as the arbiter of quality. This “vintage mythology” compresses a dynamic structural evolution into discrete temporal labels, thereby obscuring a more fundamental question: what actually transpires during aging? Addressing this question requires reframing beverage aging from time accumulation to structural maturation. In complex fluids, macroscopic properties emerge from molecular association, mesoscopic aggregation, phase-separated microdomains, and network connectivity rather than from composition alone.[1-5] Alcoholic beverages conform to this principle: mouthfeel, flavor release kinetics, and sensory persistence arise from the interplay among volatile compounds, ethanol-water clusters, non-volatile colloidal architectures, and microphase separation processes.[6-8] Aging, therefore, represents a sustained structural reorganization of the ethanol-water-flavor molecular ensemble under non-equilibrium conditions.

Baijiu provides a distinctive system for investigating such structural maturation. Its ethanol-water matrix accommodates diverse amphiphilic flavor molecules that form dynamic supramolecular networks through hydrogen bonding, hydrophobic interactions, and amphiphilic assembly.[9-13] Ethanol-water mixtures are far from ideal homogeneous solutions, exhibiting volume contraction, exothermic mixing, nanoscale heterogeneity, hydrophobic hydration, and concentration-dependent hydrogen-bond restructuring.[14-20] Yang et al. showed that ethanol-water clusters determine the critical concentration of alcoholic beverages: the wettability of ethanol-water mixtures changes stepwise with concentration, and the critical thresholds align with the standard alcohol by volume of major beverage categories.[21] This work established that empirical beverage labels can reflect intrinsic molecular cluster states. This structure-defined view of beverage properties motivates the question of whether vintage can likewise be reinterpreted as an intrinsic structural state. During Baijiu aging, liquid-liquid phase separation drives colloidal droplet shrinkage and core-shell inversion,[22] ethanol clusters evolve into dense hydrophobic lamellae that regulate flavor ester solubility,[23] and aging-induced solvent restructuring propels the system into kinetically stabilized supersaturated states.[24] These findings indicate that Baijiu aging is governed by the non-equilibrium evolution of ethanol-water-flavor molecular assemblies. Yet existing characterization methods still mainly provide local interaction, compositional, or morphological information, leaving unresolved how these structural units connect, reorganize, and mature within a unified state space.

Persistent homology provides a structural language for resolving this problem. By tracking the birth and death of topological features across a continuous filtration parameter, it converts complex point clouds or spatial structures into multiscale, noise-resilient structural fingerprints.[25-27] Topological data analysis has revealed hidden medium-range order in amorphous materials, glassy structures, polymer deformation, and disordered networks, capturing structural organization that conventional pair correlations or local descriptors often miss.[28-33] It has also been applied to defect organization, collective dynamics, and robust pattern formation in non-equilibrium soft matter and complex materials.[34-37] These studies indicate that topological invariants can transcend specific chemical identities and describe how local assemblies generate emergent order, stability, and structural transitions. For Baijiu, a chemically complex, structurally heterogeneous, and continuously evolving alcoholic soft matter system, persistent homology therefore offers a quantitative route from empirical vintage toward structural state.

Here, we applied persistent homology to alcoholic beverage aging using strong-aroma Baijiu aged from 1 to 10 years as a model system. The resulting topological evolution map revealed a three-stage maturation pathway involving early scaffold consolidation, mid-stage channel stabilization, and non-monotonic cavity reorganization during deep aging. These coupled trajectories indicated that Baijiu aging was not a featureless temporal drift, but a directional structural evolution toward a mature topological state. On this basis, we advanced the "Universal Topological Attractor" hypothesis: the optimally aged state may be defined by the position of the system in topological space relative to a mature structural domain, rather than by chronological vintage. This framework reframes aging from the accumulation of time to the navigation of structural state space, providing a physical basis for structural-fingerprint-based quality evaluation, a target state for accelerated maturation, and a general topological language for optimal self-organization in complex fluids.

# Results and Discussion

## Topological Fingerprints of Baijiu Natural Aging: A Three-Dimensional Persistent Homology Analysis

To elucidate the multiscale structural evolution occurring during Baijiu aging, three-dimensional persistent homology analysis was performed on multi-source structural characterization data obtained from strong-aroma Baijiu aged from 1 to 10 years. By tracking the birth and death of topological features across continuous filtration parameters, persistent homology transformed the colloidal assembly structures into multiscale topological fingerprints.[25,26,28,31,38] This framework simultaneously resolved the evolution of connected domains, loop-like channels, and enclosed cavities, thereby providing a global structural description of Baijiu aging beyond conventional chemical composition analysis (Figure 2A).[39]

Figures 2B and 2C summarized the global evolution of $\beta_0$, $\beta_1$, and $\beta_2$ through two complementary metrics: total feature count and average persistence. Overall, Baijiu aging exhibited a clear stagewise topological evolution rather than continuous structural relaxation. Different topological dimensions evolved asynchronously: $\beta_0$ primarily reflected the rapid establishment of a connected scaffold during early aging, $\beta_1$ captured the progressive purification and stabilization of loop structures, whereas $\beta_2$ revealed the transient generation, strengthening, and re-equilibration of enclosed cavities during middle and late aging. Together, these coordinated yet non-synchronous changes outlined a directional convergence from a fragmented heterogeneous system toward a structurally optimized mature state.

During the early aging stage (1-2 years), the dominant event involved the rapid consolidation of network connectivity. The high $\beta_0$ count observed in the 1-year sample indicated the presence of numerous isolated structural domains and a still fragmented colloidal network. By year 2, $\beta_0$ sharply collapsed into a stable low-count regime, marking the rapid emergence of a system-spanning connected scaffold. In contrast, $\beta_1$ features remained dominated by abundant short-lived rings with low average persistence, indicating that local loop structures had formed but still possessed limited structural stability. $\beta_2$ features remained scarce at this stage, suggesting that enclosed cavities and encapsulated microdomains had not yet undergone stable organization. Early aging was therefore primarily characterized by rapid skeleton formation, whereas higher-order topological structures remained highly dynamic.

From 4 to 8 years, the system entered a restructuring regime dominated by topological screening and network reorganization. $\beta_0$ counts remained largely unchanged, whereas their average persistence increased continuously, indicating progressive stabilization of the pre-established connected scaffold without large-scale network disruption. Simultaneously, $\beta_1$ counts decreased steadily, reflecting the elimination of numerous unstable short-lived rings. However, $\beta_1$ average persistence followed a non-monotonic trajectory, increasing from 8.68 at year 4 to 9.40 at year 6 before slightly decreasing to 8.92 at year 8, revealing continuous competition among ring elimination, stabilization, and network rearrangement. $\beta_2$ exhibited even stronger non-monotonicity. Its count decreased from 37 at year 4 to 25 at year 6, followed by a sharp increase to 53 at year 8, whereas the average persistence continuously increased from 0.30 to 0.70 across the same interval. The decoupling between cavity number and persistence revealed simultaneous cavity elimination, stabilization, and regeneration during middle-stage aging. Notably, the enhancement of $\beta_2$ persistence coincided with the liquid-liquid phase separation inversion previously identified by fluorescence and Raman spectroscopy,[22,40] suggesting that microphase remodeling played an important role in cavity stabilization during this period.

By year 10, the system entered a late-stage re-equilibrated state across multiple topological dimensions. $\beta_0$ remained within the low-count regime, confirming the long-term persistence of the connected scaffold established during early aging. Although $\beta_1$ counts slightly increased relative to year 8, the average $\beta_1$ persistence reached the highest value of the entire aging sequence (11.9), indicating the emergence of a globally stabilized channel network rather than continued ring elimination. In parallel, $\beta_2$ counts remained elevated, whereas the average persistence decreased from the year-8 maximum of 0.70 to 0.59, suggesting that the highly stable cavity population generated during middle aging subsequently underwent selective fusion, rearrangement, or retention. Consequently, the 10-year sample represented a mature topological state composed of a stable $\beta_0$ scaffold, highly persistent $\beta_1$ channels, and re-equilibrated $\beta_2$ cavities.

Overall, persistent homology analysis revealed that Baijiu aging represented a directional topological evolution process characterized by discrete structural reorganizations and progressive convergence toward a mature attractor state, rather than a featureless temporal drift. The following sections further dissect $\beta_0$, $\beta_1$, and $\beta_2$ individually to uncover the microstructural mechanisms underlying this three-stage maturation pathway.

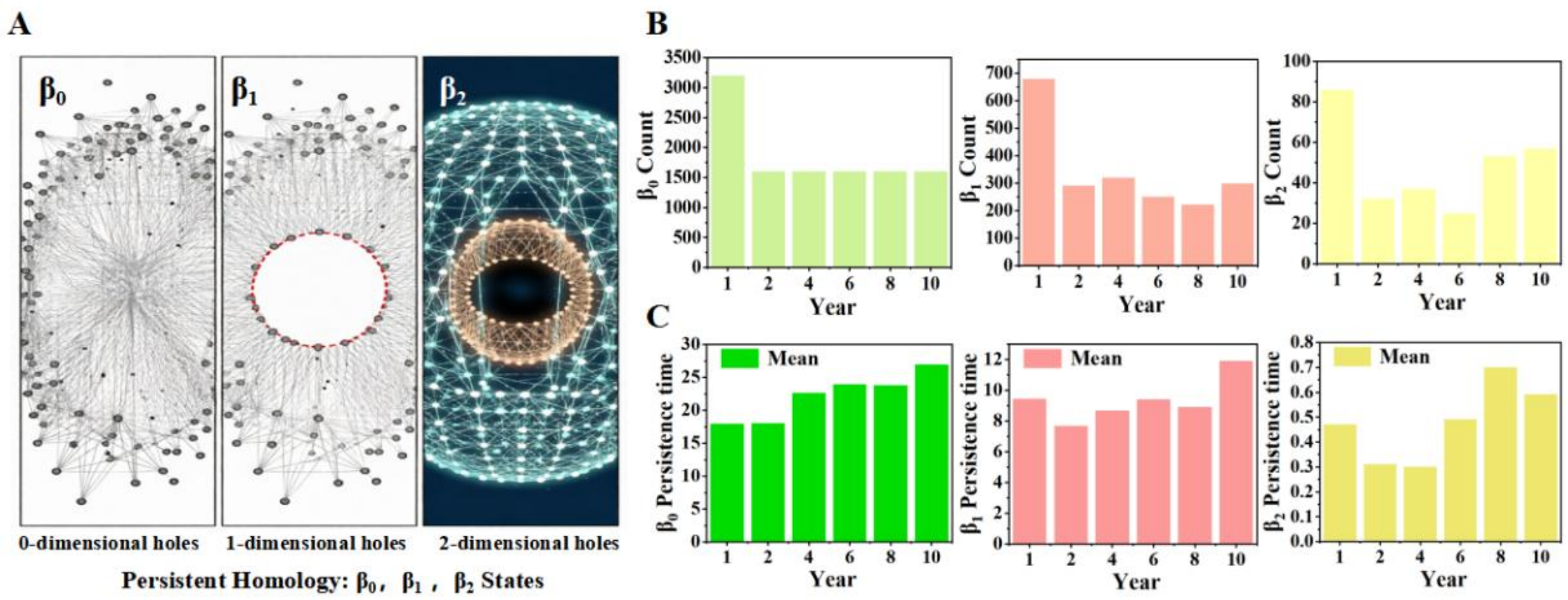

Figure 2. Global evolution of topological invariants during natural aging of Baijiu. (A) Schematic illustration of the three persistent homology invariants, $\beta_0$ (connected components), $\beta_1$ (loops), and $\beta_2$ (cavities), and their physical correspondences within a three-dimensional colloidal network. (B) Total counts of $\beta_0$, $\beta_1$, and $\beta_2$ features as functions of aging time for strong-aroma Baijiu aged 1-10 years. (C) Average persistence time of the three topological invariants across the aging series.

## $\beta_0$ (Connected Components): The Eternal Skeleton of the Colloidal Network

The zeroth homology group $\beta_0$ quantified the independent connected components within the system and served as the fundamental topological descriptor for resolving the formation of the colloidal network skeleton during Baijiu aging. In the present system, $\beta_0$ features corresponded to hydrogen-bond network fragments formed among ethanol, water, and amphiphilic flavor molecules, together with their connectivity evolution (Figure 3A). During early aging, the Baijiu system contained abundant mutually separated or weakly connected network fragments, reflecting a fragmented and structurally heterogeneous state. With progressive aging, hydrogen-bond reorganization, amphiphilic molecular association, and colloidal coalescence gradually integrated these local fragments into a stable, system-spanning connected scaffold.[41,42] Since this scaffold persisted throughout subsequent aging and provided the structural basis for the evolution of $\beta_1$ loop channels and $\beta_2$ enclosed cavities, it was termed the "eternal skeleton" of Baijiu aging.

Figure 3B quantified the evolution of $\beta_0$ persistence metrics across the aging sequence. The average $\beta_0$ persistence increased steadily from 17.93 at year 1 to 26.91 at year 10, indicating continuous enhancement in the stability and connectivity of the hydrogen-bond scaffold across the filtration scale. This increase proceeded nonlinearly, with more pronounced growth occurring during the 2-4 year and 8-10 year intervals, suggesting that skeleton stabilization involved stagewise structural reorganizations rather than uniform relaxation.[43,44] In contrast, the maximum $\beta_0$ persistence reached the saturation value of 100 by year 2 and remained unchanged thereafter. This behavior indicated that highly stable connected domains had already emerged during the initial aging stage, whereas subsequent aging primarily increased their proportion within the overall $\beta_0$ population, progressively shifting the persistence distribution toward the high-value regime.

The violin plots in Figure 3C further illustrated this population-level evolution. With increasing aging time, the $\beta_0$ persistence distribution shifted globally toward higher persistence values, while the distribution width gradually narrowed and low-persistence features diminished. The median persistence simultaneously increased from approximately 18 to approximately 24. These changes indicated that the connected scaffold evolved from a highly heterogeneous ensemble containing abundant loose local structures into a more homogeneous and collectively stabilized network. During deep aging, low-persistence connected components were progressively eliminated or integrated, whereas highly persistent domains became increasingly dominant. Consequently, the principal feature of $\beta_0$ evolution resided in the homogenization and stabilization of the connected scaffold at the population level, rather than in unlimited enhancement of individual component stability.

To further characterize this stagewise stabilization process, a topologically derived relative stability landscape was constructed from the average $\beta_0$ persistence (Figure 3D). This metric employed the reciprocal of the average $\beta_0$ persistence to represent the relative structural relaxation state of the connected scaffold and was subsequently smoothed and normalized to facilitate comparison among different aging stages. The resulting trajectory exhibited a stepwise descending pattern rather than continuous monotonic variation, indicating that $\beta_0$ evolution likely proceeded through discrete structural rearrangement events. Relatively flat intervals corresponded to metastable connected configurations, whereas rapid transitions between adjacent stages reflected cooperative reorganizations of the hydrogen-bond scaffold.[45] Together with the stagewise increase in average persistence (Figure 3B) and the progressive narrowing of the persistence distribution (Figure 3C), these results demonstrated that the $\beta_0$ skeleton evolved through sequential transitions from fragmentation to connectivity and ultimately toward homogenized stabilization. Notably, this stepwise stabilization echoed the concentration-dependent stepwise behavior of ethanol-water mixtures reported by Yang et al.,[21] where contact angles and NMR chemical shifts changed through discrete plateaus associated with tetrahedral-to-chain cluster transitions. In their system, structural thresholds emerged along the concentration axis; in the present aging system, analogous discontinuous stabilization appeared along the temporal axis, suggesting that stepwise cooperative reorganization may represent a recurring structural motif in ethanol-water-based supramolecular liquids.

Collectively, $\beta_0$ analysis revealed that the earliest and most fundamental structural event during Baijiu aging involved the rapid integration of dispersed colloidal domains into a stable connected scaffold. This skeleton was largely established during the initial aging stage and subsequently underwent progressive stabilization and population homogenization without repeated large-scale restructuring. As the foundational level of topological evolution, the $\beta_0$ scaffold provided the structural substrate for the subsequent organization and selection of higher-order $\beta_1$ and $\beta_2$ features.

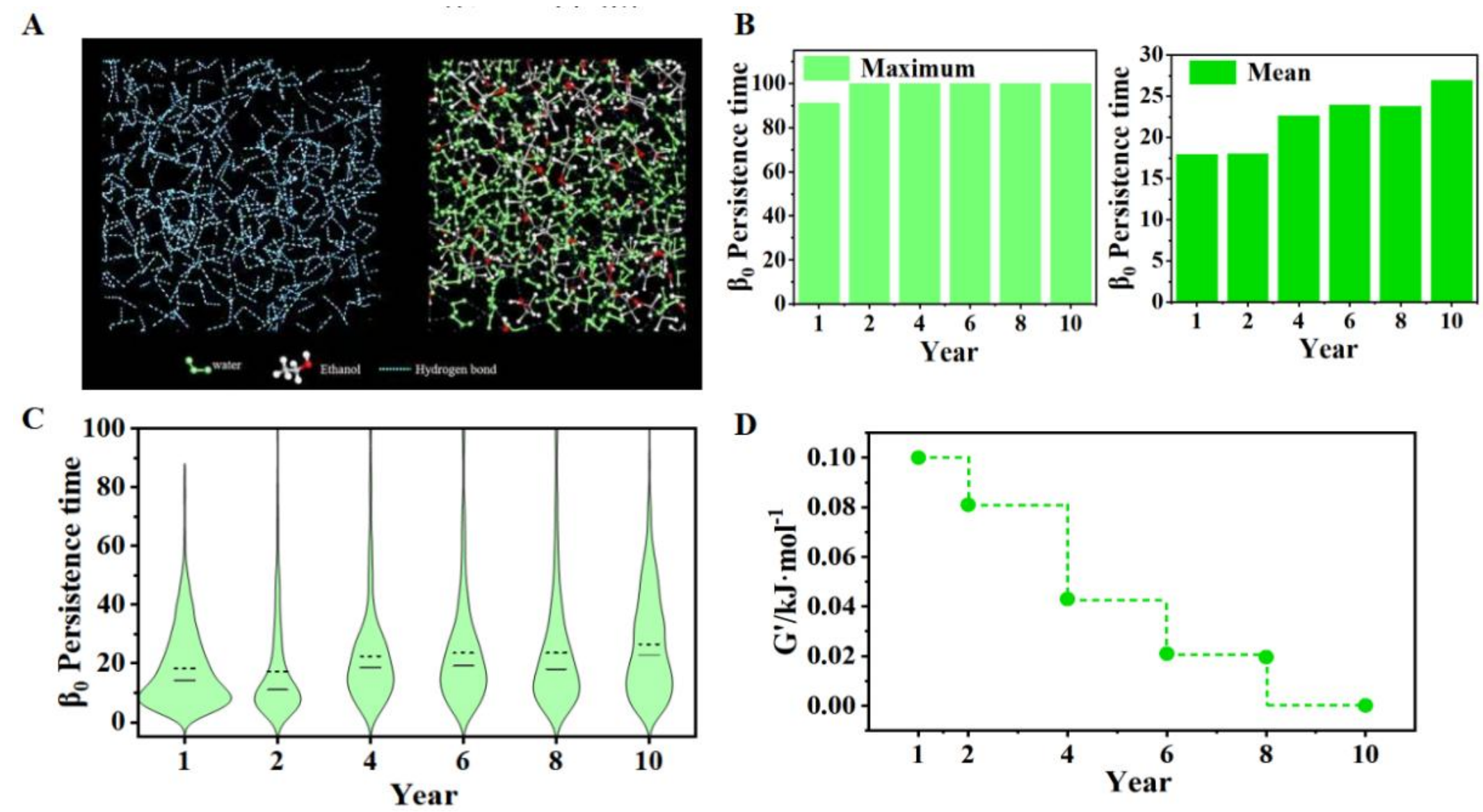

Figure 3. Evolution of $\beta_0$ connected components and consolidation of the colloidal network skeleton.

(A) Schematic of the ethanol-water hydrogen-bond network skeleton. (B) Maximum and average persistence times of $\beta_0$ as functions of aging time for strong-aroma Baijiu aged 1-10 years. (C) Violin plots of $\beta_0$ persistence time distributions for strong-aroma Baijiu aged 1-10 years, the dotted line represented the mean, and the solid line represented the median. (D) Equivalent free energy profile derived from the average $\beta_0$ persistence for strong-aroma Baijiu aged 1-10 years.

## $\beta_1$ （Loops): The Structural Heart of Aging From Disorder to Order

The first homology group $\beta_1$ characterized one-dimensional closed loops within the colloidal network and served as a key topological invariant for resolving pore and channel organization during Baijiu aging. In the present system, $\beta_1$ features represented ring-like pores or channel structures enclosed by the ethanol-water hydrogen-bond network and amphiphilic molecular assemblies. These features characterized the spatial organization of voids and cyclic connectivity within the colloidal architecture, rather than corresponding to any specific molecular channel.[46,47] Therefore, $\beta_1$ provided a quantitative descriptor for tracking how the internal pore network evolved from a dense and disordered state toward a more persistent channel architecture.

Figure 4A presented the $\beta_1$ persistence diagrams across the 1-10 year aging sequence. Early samples displayed dense point clouds near the diagonal, indicating abundant short-lived rings and a highly heterogeneous pore network. With increasing aging time, the diagonal population gradually diminished, whereas features farther from the diagonal became more prominent. This redistribution indicated that aging progressively eliminated fragile local pores and enriched longer-lived loop structures. The $\beta_1$ evolution therefore reflected a population-level purification of the ring network, rather than the continuous emergence of individual rings with ever-increasing maximum persistence.

The kernel density estimation and violin plots further confirmed this distributional shift (Figures 4B and 4C). The 1-year sample was dominated by low-persistence $\beta_1$ features, whereas aging progressively shifted the distribution toward intermediate and high persistence regimes. From year 6 to year 8, a high-persistence shoulder emerged, indicating the accumulation of long-lived ring features within the population. By year 10, the low-persistence contribution had markedly decreased, and intermediate-to-high persistence features became dominant. Consistently, the violin plots showed an upward extension of the $\beta_1$ persistence distribution and a gradual reduction of the low-persistence population. These results demonstrated that $\beta_1$ maturation was governed by statistical redistribution of the entire loop population, in which unstable rings were removed or reorganized while more persistent channel-like structures became increasingly enriched.

Figure 4D quantified this process using maximum and average $\beta_1$ persistence. The maximum $\beta_1$ persistence followed a non-monotonic trajectory, decreasing from 57.18 at year 1 to 41.33 at year 4, recovering to 46.09 at year 6, decreasing again to 41.35 at year 8, and reaching 47.51 at year 10. After year 4, the maximum persistence remained confined within a relatively narrow range of 41-47, indicating that the upper stability limit of individual rings did not continuously increase with aging. In contrast, the average $\beta_1$ persistence showed a clearer population-level increase: after a transient decline from 9.43 at year 1 to 7.66 at year 2, it rose to 8.68 at year 4 and 9.40 at year 6, slightly decreased to 8.92 at year 8, and reached the highest value of 11.90 at year 10. This divergence between maximum and average persistence indicated that late-stage aging primarily increased the proportion of stable ring structures within the $\beta_1$ population, rather than generating isolated extreme rings.

The transient decrease in average $\beta_1$ persistence at year 2 likely reflected the transitional restructuring of the pore network during early scaffold consolidation. The rapid formation of the $\beta_0$ connected skeleton could have merged, compressed, or reorganized preexisting local loops, whereas stable channel structures had not yet fully developed. After this transition, loop structures were progressively re-screened and stabilized on the established $\beta_0$ scaffold. From year 2 to year 10, the average $\beta_1$ persistence increased by approximately 1.55-fold, demonstrating the progressive population-level stabilization of the channel network during aging.

Above all, $\beta_1$ analysis revealed a topological purification process in which the early disordered pore network was remodeled into a more stable and coherent channel architecture. The defining feature of this process was not the continuous increase of single-ring maximum persistence, but the gradual elimination of short-lived loops and enrichment of long-lived loop populations. Built upon the $\beta_0$ connected skeleton, this stabilized $\beta_1$ channel network may provide a more ordered spatial environment for the confinement, diffusion, and sustained release of flavor molecules, thereby offering a potential structural basis for the smooth and harmonious sensory characteristics of aged Baijiu.[48,49]

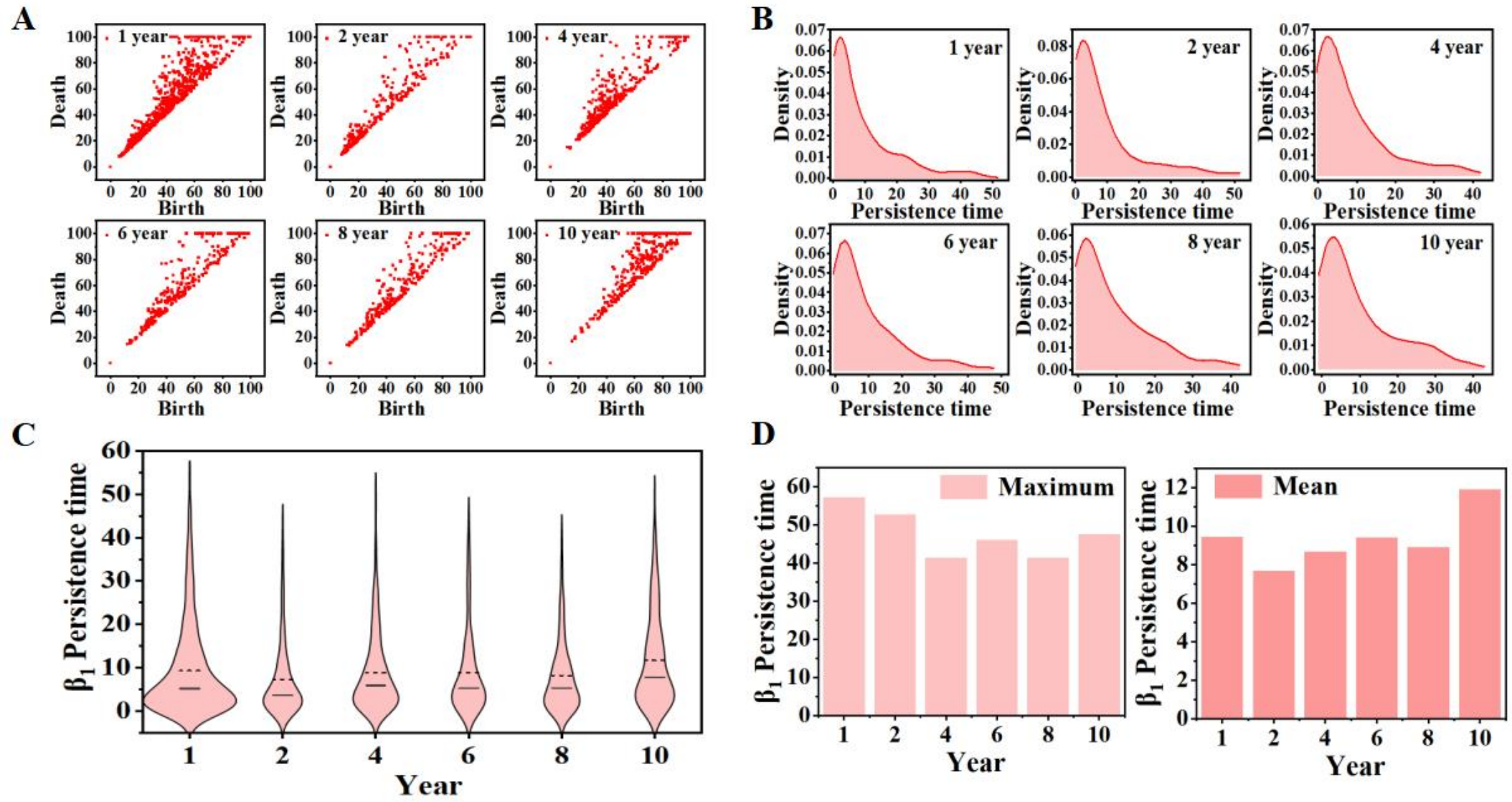

Figure 4. Topological purification and stabilization evolution of the $\beta_1$ loop network. (A) $\beta_1$ persistence diagrams for strong-aroma Baijiu aged 1-10 years. (B) Kernel density estimation curves of $\beta_1$ feature persistence times for strong-aroma Baijiu aged 1-10 years. (C) Violin plots of $\beta_1$ persistence time distributions for strong-aroma Baijiu aged 1-10 years, the dotted line represented the mean, and the solid line represented the median. (D) Maximum and average persistence times of $\beta_1$ as functions of aging time for strong-aroma Baijiu aged 1-10 years.

## $\beta_2$ (Cavities): The Topological Signature of Deep Aging

The second homology group $\beta_2$ described fully enclosed cavities in three-dimensional space and served as a topological descriptor of higher-order spatial confinement. In the Baijiu colloidal system, $\beta_2$ features corresponded to enclosed microdomains or vesicle-like cavities formed by the ethanol-water hydrogen-bond network and amphiphilic flavor molecules. In contrast to the $\beta_0$ connected scaffold and $\beta_1$ loop channels, $\beta_2$ captured encapsulated structures whose emergence, disappearance, and reorganization reflected changes in liquid-liquid phase separation, hydrophobic microdomain formation, and internal spatial confinement during aging.[50,51] Thus, $\beta_2$ provided a sensitive topological marker for mid-to-late structural maturation.

Figure 5A presented the evolution of $\beta_2$ persistence distributions across the 1-10 year aging sequence. Unlike $\beta_1$, which mainly exhibited population-level stabilization, $\beta_2$ followed a markedly non-monotonic pathway. The 1-year sample already contained a heterogeneous population of enclosed microdomains, whereas the 2-year sample showed a contracted persistence distribution, indicating the removal or rearrangement of early metastable cavities during scaffold consolidation. The $\beta_2$ population expanded again after year 4 and extended toward higher persistence at year 6, revealing the transient formation of highly stable enclosed cavities. By year 8, high-persistence $\beta_2$ features became more broadly represented within the population, indicating that cavity stabilization had expanded from isolated extreme structures to a larger set of enclosed microdomains. At year 10, the distribution shifted toward a more balanced state, suggesting selective retention and re-equilibration rather than continued amplification of all high-persistence cavities.

Figure 5B quantified this non-monotonic behavior using maximum and average $\beta_2$ persistence. The maximum $\beta_2$ persistence decreased from 1.85 at year 1 to 0.64 at year 4, surged to the series maximum of 2.87 at year 6, and then decreased to 2.68 at year 8 and 2.19 at year 10. This trajectory identified year 6 as the point at which the most stable individual cavity emerged. By contrast, the average $\beta_2$ persistence peaked at year 8, increasing from 0.30 at year 4 to 0.49 at year 6 and 0.70 at year 8 before decreasing to 0.59 at year 10. The temporal offset between the maxima of individual and average persistence defined the central feature of $\beta_2$ evolution: year 6 represented the transient generation of exceptional cavities, whereas year 8 represented the population-level stabilization of enclosed microdomains.

Figure 5C highlighted the representative high-persistence $\beta_2$ feature that emerged specifically in the 6-year sample. This cavity exhibited a birth value of 6.05, a death value of 8.92, and a persistence of 2.87, corresponding to the maximum $\beta_2$ persistence observed across the entire aging series. Combined with previous fluorescence and Raman evidence for liquid-liquid phase separation inversion, this feature suggested that year 6 represented a critical window of phase-separated structural remodeling.[22] During this stage, reorganization of local hydrogen-bond networks and hydrophobic microdomains likely enabled the transient formation of highly stable encapsulated structures. The absence of this extreme cavity in the 8- and 10-year samples indicated that it represented an intermediate state rather than a terminal mature configuration. Subsequent aging likely redistributed such cavities through fusion, collapse, or incorporation into larger assemblies, leading to a more balanced $\beta_2$ population.[52,53]

Taken together, $\beta_2$ analysis revealed a cavity reorganization pathway distinct from the early consolidation of $\beta_0$ and the population-level stabilization of $\beta_1$. The pathway involved the clearance of early metastable cavities, the transient emergence of highly stable enclosed microdomains, the subsequent enrichment of stable cavities at the population level, and final re-equilibration during deep aging. This sequence established $\beta_2$ as an important topological signature of mid-to-late Baijiu maturation. The formation and stabilization of enclosed microdomains may provide spatial environments for the confinement and sustained release of hydrophobic flavor molecules, thereby offering a structural framework for interpreting flavor persistence and prolonged aftertaste in aged Baijiu.[54,55]

Figure 5D further integrated the three topological maturation milestones with sensory quality development. At year 2, the rapid stabilization of $\beta_0$ indicated the formation of a structurally integrated hydrogen-bond scaffold, corresponding to the transition from a fragmented to a globally connected state and potentially to the initial attenuation of pungency. At year 8, the appearance of a representative ultra-stable $\beta_1$ ring indicated that the channel architecture had entered a highly stabilized regime, which may support more ordered confinement, diffusion, and balanced release of flavor molecules and thus relate to improved smoothness and harmony. Between years 6 and 8, $\beta_2$ evolved from the transient formation of exceptional cavities to the population-level stabilization of enclosed microdomains, suggesting a possible structural basis for hydrophobic flavor encapsulation, sustained release, and lingering aftertaste. This topology-sensory correspondence was conceptually consistent with the temperature-dependent cluster transitions reported by Yang et al., in which ethanol-water cluster restructuring was linked to altered “ethanol-like” taste and preferred drinking temperature.[21] Their work provided a parallel example of how sensory perception can be modulated by molecular-scale structural states. Our present results extended this structure-perception principle to aging, where $\beta_0$ scaffold integrity, $\beta_1$ channel stabilization, and $\beta_2$ cavity reorganization provided a topological framework for interpreting pungency attenuation, smoothness enhancement, and prolonged aftertaste.

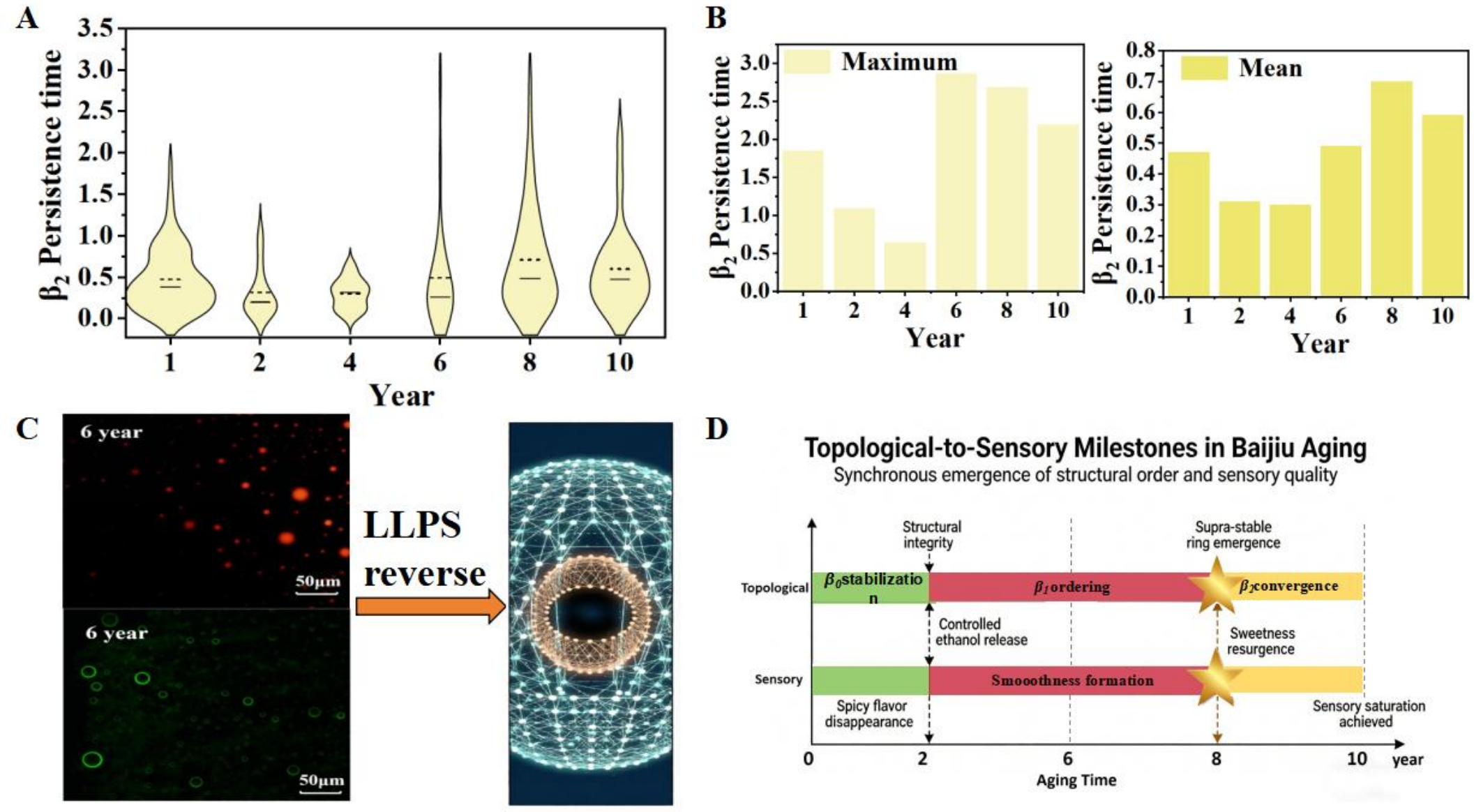

Figure 5. Non-monotonic evolution of $\beta_2$ cavities and the topological-sensory correlation during deep aging.

(A) Violin plots of $\beta_2$ feature persistence time distributions for strong-aroma Baijiu aged 1-10 years, the dotted line represented the mean, and the solid line represented the median. (B) Maximum and average persistence times of $\beta_2$ as functions of aging time for strong-aroma Baijiu aged 1-10 years. (C) Fluorescence staining image of 6-year Baijiu and a schematic of the corresponding $\beta_2$ cavity. (D) Schematic diagram illustrating the correspondence between topological maturation milestones and sensory quality enhancement.

## The Universal Topological Attractor: From Empirical Vintage to Structural Physics

The preceding analyses revealed that strong-aroma Baijiu aged from 1 to 10 years underwent staged topological reorganization across connected scaffolds, loop channels, and enclosed cavities. Although $\beta_0$, $\beta_1$, and $\beta_2$ followed distinct temporal trajectories, their coordinated evolution supported a unified picture: Baijiu aging represented a directional convergence toward a mature structural state, rather than a passive accumulation of chronological time.[56]

Figure 6A projected samples of different aging years into a three-dimensional topological space jointly defined by $\beta_0$, $\beta_1$, and $\beta_2$. The samples followed a directional trajectory rather than a random distribution, indicating that natural aging proceeded along a well-defined structural pathway. Within this space, the 8-year sample occupied a low-energy region associated with $\beta_2$ cavity stabilization and the emergence of highly stable $\beta_1$ ring structures, suggesting a metastable intermediate state during mid-to-late aging.[57] In contrast, the 10-year sample exhibited a stabilized $\beta_0$ scaffold, the highest average $\beta_1$ persistence across the series, and a re-equilibrated $\beta_2$ cavity population after the transient high-stability stage. Thus, year 8 represented a low-energy intermediate associated with cavitation and channel stabilization, whereas year 10 corresponded to a mature attractor state arising from coordinated stabilization across multiple topological levels.

Based on this trajectory, we proposed the "Universal Topological Attractor" hypothesis (Figure 6B): the optimally aged state of an alcoholic beverage may be defined by its position in topological space relative to a mature structural domain, rather than by absolute vintage years.[58,59] In this framework, the ethanol-water matrix provides a common structural background, while trace flavor molecules modulate local topology through hydrogen bonding, hydrophobic interactions, and amphiphilic assembly. Different aroma types, alcohol contents, or aging pathways may follow distinct kinetic routes, yet their optimized states may approach attractor regions with similar topological characteristics. Vintage therefore functions as an external temporal label, whereas the topological fingerprint provides an intrinsic coordinate for structural maturity.

This hypothesis yields several testable predictions. Optimally aged Baijiu of different aroma types should approach similar mature topological regions rather than vary monotonically with chronological age. Accelerated aging should be considered structurally equivalent only if it drives samples toward the same attractor region as natural aging. Over-aging should appear as deviation from the mature attractor, rather than as continued enhancement of topological metrics. Finally, the topological distance to the attractor should correlate with sensory quality, flavor release behavior, and colloidal stability.

Persistent homology therefore provided a new structural basis for redefining vintage in Baijiu aging. Traditional evaluation treats time as the primary measure of quality; the topological attractor framework instead defines aging as the convergence of a complex fluid toward a specific structural state under non-equilibrium conditions. This framework may shift quality assessment from empirical vintage judgment to structural-fingerprint-based characterization, and may redefine accelerated aging as the controlled navigation of topological space rather than the simple compression of time. More broadly, the Universal Topological Attractor hypothesis offers a quantifiable structural language for understanding optimal self-organization in alcoholic beverages and other complex fluids.

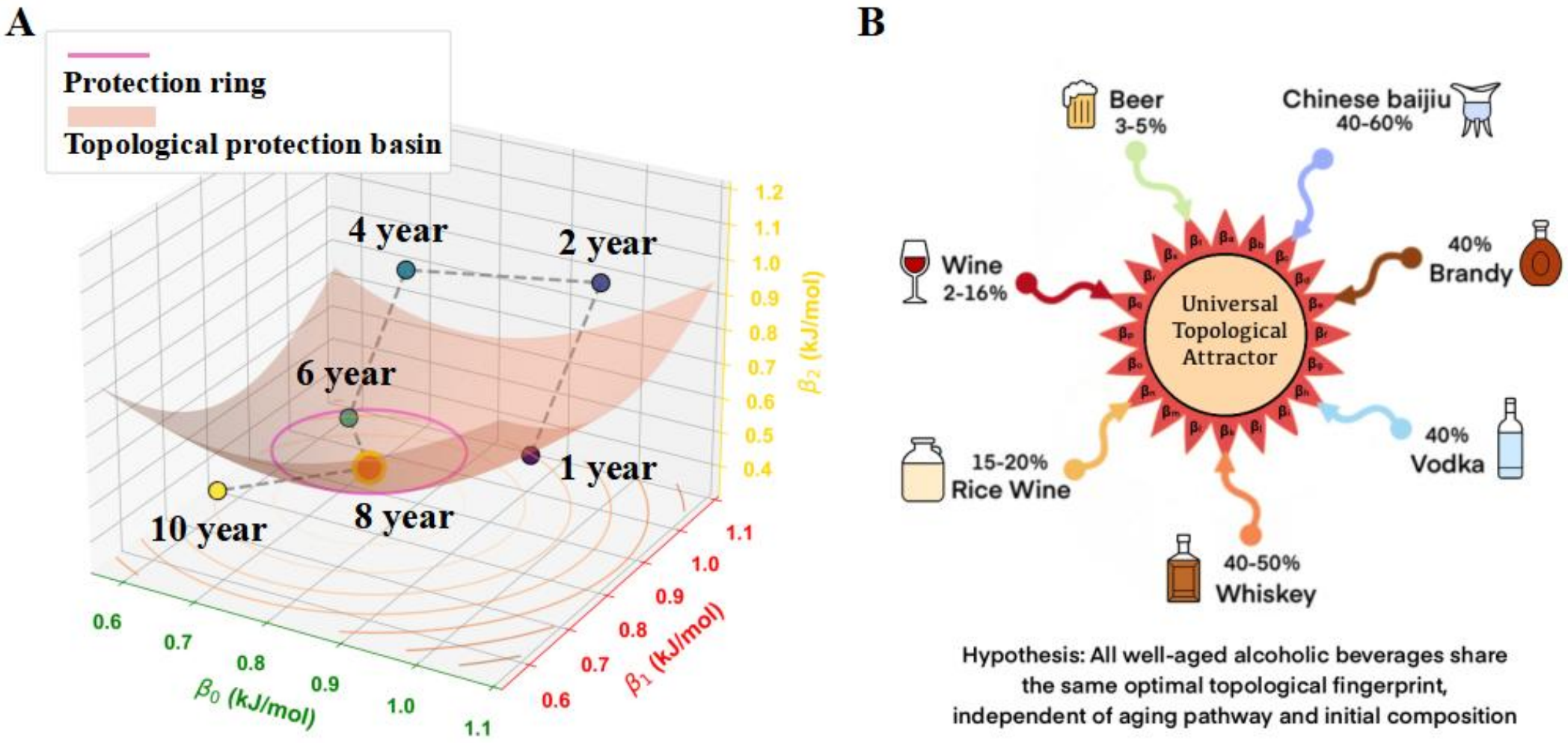

Figure 6. Universal topological attractor and the paradigm shift from vintage to structure.
(A) Three-dimensional topological-energy scatter plot constructed from the average persistence times of $\beta_0$, $\beta_1$, and $\beta_2$ via free energy scaling for strong-aroma Baijiu aged 1 to 10 years. (B) Schematic of the Universal Topological Attractor hypothesis.

# CONCLUSION

This study introduced persistent homology as a topological framework for decoding the structural maturation of Baijiu aging. Using strong-aroma Baijiu aged from 1 to 10 years, we constructed a three-dimensional topological evolution map of self-assembled molecular aggregates and revealed that aging proceeded as a staged, directional, and non-monotonic structural evolution rather than a simple accumulation of time. $\beta_0$ captured the rapid establishment of a connected hydrogen-bond skeleton within the early aging stage; $\beta_1$ revealed the population-level stabilization of loop-like channels during maturation; and $\beta_2$ identified a non-monotonic cavity reorganization process associated with mid-to-late spatial confinement. Together, these topological trajectories demonstrated that Baijiu aging was governed by the coordinated evolution of connected scaffolds, loop channels, and enclosed cavities toward a mature structural state. Based on this convergence, we proposed the "Universal Topological Attractor" hypothesis, in which optimal aging may be defined by the position of a system in topological space relative to a mature structural domain rather than by chronological vintage. Building on the structure-defined concentration paradigm established by Yang et al., this work extends the structure-property principle of alcoholic beverages from static ethanol-water cluster thresholds to dynamic topological maturation. This perspective indicates that beverage quality should be understood not through external empirical labels such as ABV or vintage, but through intrinsic structural states. This framework provides a structural-physics basis for redefining beverage aging, guiding accelerated maturation, and understanding optimal self-organization in complex fluids.

# EXPERIMENTAL PROCEDURES

## Baijiu samples

The Baijiu samples were collected from the manufacturer Jing Brand Co. Ltd., Daye, China. Baijiu samples were aged for 6 different years (1- year aged, 2-year aged, 4-year aged, 6-year aged, 8-year aged, and 10- year aged Baijiu). The alcohol content of Baijiu samples was 68 % (v/ v). All Baijiu samples used in this study were directly collected from storage containers without any additives.

## Fluorescence staining imaging

In the single staining experiment, 6 nM fluorescein and Rhodamine B were added into Baijiu samples respectively, mixed evenly and placed away from light for 24 h. After that, 6 ~ 8 μL single staining Baijiu samples were dropped on the bottom glass and covered with a cover glass for fluorescence microscope observation. The fluorescent signal of fluorescein is green, and that of Rhodamine B is orange. In the double-staining experiment, 6nM fluorescein and Rhodamine B were added into Baijiu samples, and the subsequent operations were similar.

Confocal fluorescence microscopy observations were carried out using a Nikon D-Eclipse C1 controller combined with a EC-C1 software. The instrument has also been connected to a Nikon eclipse TE 2000-U fluorescence microscope with an inverted lens. This allows us to preview the sample before confocal measurements. For samples labeled with fluorescein, a He/Ne laser with a wavelength of 485 nm was used. For samples labeled with Rhodamine B, an argon laser with a wavelength of 545 nm was selected. For samples labeled with mixed stains, both lasers were applied and the intensity of individual laser was adjusted in a way that the signals of orange color and green color are comparable. The EC-C1 software allows us to obtain micrographs with both mixed color and individual color by opening or closing each color channel.

## Raman spectroscopy

A confocal Raman microscope (XploRA Plus, HORIBA, Japan) equipped with a 532 nm laser (10% laser intensity), 1800 gr/mm grating, and a 50 × (LWD visible objective, NA=0.50, WD=10.6 mm) magnification objective lens was used for Raman analysis. Data were collected with an integration time of 1 s for measurements of single spectrum, or 0.5 s for image mapping. For Raman image mapping, the sample was scanned over an area of 10 μm × 10 μm using a 50× objective lens, or 50 μm × 50 μm using a 100× objective lens. The methyl characteristic peak at 2926.8 cm-1 was used for image mapping to obtain the whole picture of Baijiu droplet distribution.

## Electrochemical measurements

All electrochemical experiments were conducted using a CorrTest electrochemical workstation and a glass three-electrode cell, with Pt wire serving as the counter electrode and reference electrode. The working electrode employed was the GUITAR electrode utilized in previous study.[60] A sinusoidal voltage ranging from 0.1 M Hz to 10 Hz was applied to the Baijiu samples. A sampling rate of 10 frequency points per decade was adopted. To ensure temperature stability, a thermostatic water bath maintained the measuring system at 25 ℃ ± 0.1 ℃.

## Enzymatic hydrolysis fluorescence intensity

The ADH activity was determined according to the protocol reported by Lee et al., with slight modifications. The following solutions were prepared: 300 μL of 3 M Tris-HCl (adjusted to pH 8.0 at 25℃), 300 μL of freshly prepared 2.5 mg/L β-$NAD^+$, 300 μL of 3 U/mL ADH and 50 μL of substrate (Baijiu sample or ethanol-water binary solution). The activity of ADH was assessed by measuring the formation of NADH, using an ELISA plate reader (BioTek ELx808, Mumbai, India) at 340 nm. The reaction proceeded at 25℃ and pH 8.0. A control reaction was conducted in the absence of any test sample. A blank reaction (in the absence of the substrate) was also measured, and values were baseline-corrected to eliminate the effect of any artifacts.

## Steady-state fluorescence spectroscopy

Measurements were conducted using an F-7000 steady-state fluorescence spectrophotometer (Hitachi High-Technologies Corporation, Japan).

Blank Baijiu Sample Analysis: A 3 mL aliquot of untreated Baijiu was transferred to a 1 cm quartz cuvette. Steady-state fluorescence spectra were recorded at 25 ± 1 ◦C with the excitation wavelength set at 255 nm.

Fluorescent probe measurement: Fluorescein and Rhodamine B dyes were accurately weighed and dissolved in chromatographic grade ethanol to prepare 0.03 mmol/L stock solutions. Coumarin, Rhodamine B, and Nile Red stock solutions were separately added to 5 mL Baijiu samples to adjust their final concentrations to 60 nmol/L. The mixtures were magnetically stirred for 10 min and then allowed to stand for equilibration for 30 min. Fluorescence spectra were measured using the same instrument parameters as mentioned above. Specifically, the excitation wavelength was set to 485 nm for fluorescein detection and 545 nm for Rhodamine B detection. The emission spectra and changes in fluorescence intensity were recorded simultaneously. Each sample was measured in triplicate, and the average value was used for data analysis.

## Persistent homology analysis and vectorization

Persistent homology (PH) was used to extract topological features from all four types of experimental data. The computations were performed using the GUDHI library (v3.8.0) in Python 3.9. The general workflow consisted of: (i) converting each data modality into a point cloud in $R^2$ or $R^3$; (ii) constructing a Vietoris-Rips complex with a maximum edge length equal to 0.1× the diagonal length of the point cloud bounding box; (iii) computing persistence intervals for dimensions 0 (connected components), 1 (loops), and, for 3D point clouds, dimension 2 (voids); (iv) filtering out intervals with a lifetime (death - birth) smaller than 0.05× the maximum edge length; and (v) exporting the (birth, death) pairs for each dimension.

Fluorescence images (2D grayscale): each pixel with non-zero intensity was converted to a point (x, y, I), where x and y are the pixel coordinates and I is the intensity value. The intensity was scaled by max(width, height)/255 to match the spatial coordinate scale. The resulting 3D point cloud was then subjected to PH up to dimension 2.

Raman spectra, enzymatic fluorescence intensity curves, and steady-state fluorescence spectra: each spectrum (intensity vs. wavenumber or time) was downsampled by a factor of 5 to reduce redundant points and then mapped to a 2D point set (x = index, y = normalized intensity). The intensity axis was standardized using StandardScaler. PH was computed only for dimensions 0 and 1.

EIS data: the real and imaginary parts of impedance formed a 2D point cloud (Z′, -Z″). The points were standardized to zero mean and unit variance before PH computation (dimensions 0 and 1).

For all point clouds, if the number of points exceeded 2000, a random subset of 2000 points was selected with a fixed random seed (42) to ensure reproducibility. The computed (birth, death) pairs for each dimension were saved as separate Excel files for subsequent analysis.

# AUTHOR CONTRIBUTIONS

Conceptualization, Y.X., and Y.L.; Methodology, Y.X., Y.L., and X.J.; Investigation, X.J., H.Y., Z.J. and L.C.; Writing-original draft, X.J. and H.Y.; writing-review & editing, Y.X., and Y.L.; Resources, Y.X., and Y.L.; Supervision, Y.X.

# DECLARATION OF INTERESTS

The authors declare no competing interests.

# REFERENCES

1. Cipelletti, L., and Ramos, L. (2005). Slow dynamics in glassy soft matter. J. Phys.: Condens. Matter *17*, R253. https://doi.org/10.1088/0953-8984/17/6/R01.
2. Trappe, V., Prasad, V., Cipelletti, L., Segre, P.N., and Weitz, D.A. (2001). Jamming phase diagram for attractive particles. Nature *411*, 772-775. https://doi.org/10.1038/35081021.
3. Lu, P.J., Zaccarelli, E., Ciulla, F., Schofield, A.B., Sciortino, F., and Weitz, D.A. (2008). Gelation of particles with short-range attraction. Nature *453*, 499-503. https://doi.org/10.1038/nature06931.
4. Royall, C.P., Faers, M.A., Fussell, S.L., and Hallett, J.E. (2021). Real space analysis of colloidal gels: triumphs, challenges and future directions. J. Phys.: Condens. Matter *33*, 453002. https://doi.org/10.1088/1361-648X/ac04cb.

5. Bonn, D., Denn, M.M., Berthier, L., Divoux, T., and Manneville, S. (2017). Yield stress materials in soft condensed matter. Rev. Mod. Phys. *89*, 035005. https://doi.org/10.1103/RevModPhys.89.035005.
6. Ammari, A., and Schroen, K. (2018). Flavor Retention and Release from Beverages: A Kinetic and Thermodynamic Perspective. Journal of Agricultural and Food Chemistry *66*, 9869-9881. https://doi.org/10.1021/acs.jafc.8b04459.
7. Ickes, C.M., and Cadwallader, K.R. (2017). Effects of Ethanol on Flavor Perception in Alcoholic Beverages. Chemosensory Perception *10*, 119-134. https://doi.org/10.1007/s12078-017-9238-2.
8. Ickes, C.M., and Cadwallader, K.R. (2018). Effect of ethanol on flavor perception of Rum. Food Science & Nutrition *6*, 912-924. https://doi.org/10.1002/fsn3.629.
9. Jiang, X., Liu, D., Yang, S., Cheng, X., and Xie, Y. (2024). Evolution of self-assembled amphiphilic colloidal particles in strong-flavor Chinese baijiu. Food Chem. *461*, 140883. https://doi.org/10.1016/j.foodchem.2024.140883.
10. Jiang, X., Liu, R., and Xie, Y. (2024). Hydrogen bonding dominated self-assembly mechanism of amphiphilic molecules in Chinese Baijiu. Food Chem. *452*, 139420. https://doi.org/10.1016/j.foodchem.2024.139420.
11. Xu, Y., Zhao, J., Liu, X., Zhang, C., Zhao, Z., Li, X., and Sun, B. (2022). Flavor mystery of Chinese traditional fermented baijiu: The great contribution of ester compounds. Food Chem. *369*, 130920. https://doi.org/10.1016/j.foodchem.2021.130920.
12. Qiao, L., Wang, J., Wang, R., Zhang, N., and Zheng, F. (2023). A review on flavor of Baijiu and other world-renowned distilled liquors. Food Chemistry: X *20*, 100870. https://doi.org/10.1016/j.fochx.2023.100870.
13. Ma, X., Sun, Y., Pan, D., Cao, J., and Dang, Y. (2022). Structural characterization and stability analysis of phosphorylated nitrosohemoglobin. Food Chem. *373*, 131475. https://doi.org/10.1016/j.foodchem.2021.131475.
14. Franks, F., and Ives, D.J.G. (1966). The structural properties of alcohol–water mixtures. Quarterly Reviews, Chemical Society *20*, 1-44. https://doi.org/10.1039/QR9662000001.
15. Parke, S.A., and Birch, G.G. (1999). Solution properties of ethanol in water. Food Chem. *67*, 241-246. https://doi.org/10.1016/S0308-8146(99)00124-7.
16. Gereben, O., and Pusztai, L. (2015). Investigation of the Structure of Ethanol–Water Mixtures by Molecular Dynamics Simulation I: Analyses Concerning the Hydrogen-Bonded Pairs. The Journal of Physical Chemistry B *119*, 3070-3084. https://doi.org/10.1021/jp510490y.
17. Pothoczki, S., Pethes, I., Pusztai, L., Temleitner, L., Ohara, K., and Bakó, I. (2021). Properties of Hydrogen-Bonded Networks in Ethanol–Water Liquid Mixtures as a Function of Temperature: Diffraction Experiments and Computer Simulations. The Journal of Physical Chemistry B *125*, 6272-6279. https://doi.org/10.1021/acs.jpcb.1c03122.
18. Dolenko, T.A., Burikov, S.A., Dolenko, S.A., Efitorov, A.O., Plastinin, I.V., Yuzhakov, V.I., and Patsaeva, S.V. (2015). Raman Spectroscopy of Water–Ethanol Solutions: The Estimation of Hydrogen Bonding Energy and the Appearance of Clathrate-like Structures in Solutions. The Journal of Physical Chemistry A *119*, 10806-10815. https://doi.org/10.1021/acs.jpca.5b06678.

19. Qin, D., Shen, Y., Yang, S., Zhang, G., Wang, D., Li, H., and Sun, J. (2022). Whether the Research on Ethanol–Water Microstructure in Traditional Baijiu Should Be Strengthened? Molecules *27*, 8290. https://doi.org/10.3390/molecules27238290.
20. Shang, Y., Hajar, R., Jiang, X., and Xie, Y. (2025). Unraveling the Anomalous Physicochemical Properties of Ethanol–Water Binary Solutions via Hydrogen-Bond-Driven Self-Assembly of Ethanol Clusters. The Journal of Physical Chemistry A *129*, 6837-6844. https://doi.org/10.1021/acs.jpca.5c03699.
21. Yang, X., Zheng, J., Luo, X., Xiao, H., Li, P., Luo, X., Tian, Y., Jiang, L., and Zhao, D. (2024). Ethanol-water clusters determine the critical concentration of alcoholic beverages. Matter *7*, 1724-1735. https://doi.org/10.1016/j.matt.2024.03.017.
22. Jiang, X., Shang, Y., Hajar, R., Yang, H., Peng, J., Li, J., Wang, W., Zou, Z., and Xie, Y. (2026). Evolutionary pattern of liquid-liquid phase separation in amphiphilic molecular self-assembly during the natural aging process of strong-aroma Baijiu. Food Res. Int. *225*, 118060. https://doi.org/10.1016/j.foodres.2025.118060.
23. Shang, Y., Jiang, X., Zuo, Y., and Xie, Y. (2026). Dynamic evolution of ethanol clusters and solubility modulation of flavor esters during the aging of Soy sauce flavor Baijiu. Food Chem. *505*, 148095. https://doi.org/10.1016/j.foodchem.2026.148095.
24. Zhao, C., Jiang, X., and Xie, Y. (2026). Liquid–liquid phase separation and self-assembly of hexanoic acid and ethyl hexanoate in ethanol–water systems: a model for aged colloidal Baijiu. Soft Matter *22*, 2958-2966. https://doi.org/10.1039/D6SM00143B.
25. Edelsbrunner, Letscher, and Zomorodian (2002). Topological Persistence and Simplification. Discrete & Computational Geometry *28*, 511-533. https://doi.org/10.1007/s00454-002-2885-2.
26. Carlsson, and Gunnar (2009). TOPOLOGY AND DATA. Bulletin of the American Mathematical Society. https://doi.org/10.1090/S0273-0979-09-01249-X.
27. Ghrist, R. (2008). Barcodes: The persistent topology of data. Bulletin of the American Mathematical Society *45*, 61-75. https://doi.org/10.1090/S0273-0979-07-01191-3.
28. Hiraoka, Y., Nakamura, T., Hirata, A., Escolar, E.G., Matsue, K., and Nishiura, Y. (2016). Hierarchical structures of amorphous solids characterized by persistent homology. Proceedings of the National Academy of Sciences *113*, 7035-7040. https://doi.org/10.1073/pnas.1520877113.

29. Nakamura, T., Hiraoka, Y., Hirata, A., Escolar, E.G., and Nishiura, Y. (2015). Persistent homology and many-body atomic structure for medium-range order in the glass. Nanotechnology *26*, 304001. https://doi.org/10.1088/0957-4484/26/30/304001.
30. Ichinomiya, T., Obayashi, I., and Hiraoka, Y. (2017). Persistent homology analysis of craze formation. Physical Review E *95*, 012504. https://doi.org/10.1103/PhysRevE.95.012504.
31. Sørensen, S.S., Biscio, C.A.N., Bauchy, M., Fajstrup, L., and Smedskjaer, M.M. (2020). Revealing hidden medium-range order in amorphous materials using topological data analysis. Science Advances *6*, eabc2320. https://doi.org/10.1126/sciadv.abc2320.
32. Sørensen, S.S., Du, T., Biscio, C.A.N., Fajstrup, L., and Smedskjaer, M.M. (2022). Persistent homology: A tool to understand medium-range order glass structure. Journal of Non-Crystalline Solids: X *16*, 100123. https://doi.org/10.1016/j.nocx.2022.100123.
33. Obayashi, I., Nakamura, T., and Hiraoka, Y. (2022). Persistent Homology Analysis for Materials Research and Persistent Homology Software: HomCloud. J. Phys. Soc. Jpn. *91*, 091013. https://doi.org/10.7566/JPSJ.91.091013.
34. Saw, T.B., Doostmohammadi, A., Nier, V., Kocgozlu, L., Thampi, S., Toyama, Y., Marcq, P., Lim, C.T., Yeomans, J.M., and Ladoux, B. (2017). Topological defects in epithelia govern cell death and extrusion. Nature *544*, 212-216. https://doi.org/10.1038/nature21718.
35. Shankar, S., Scharrer, L.V.D., Bowick, M.J., and Marchetti, M.C. (2024). Design rules for controlling active topological defects. Proceedings of the National Academy of Sciences *121*, e2400933121. https://doi.org/10.1073/pnas.2400933121.
36. Zheng, Z., Jiang, C., Chen, Y., Baggioli, M., and Zhang, J. (2026). Topological signatures of collective dynamics and turbulent-like energy cascades in apolar active granular matter. Proceedings of the National Academy of Sciences *123*, e2510873123. https://doi.org/10.1073/pnas.2510873123.
37. Tokura, Y., and Kanazawa, N. (2021). Magnetic Skyrmion Materials. Chem. Rev. *121*, 2857-2897. https://doi.org/10.1021/acs.chemrev.0c00297.
38. Cramer Pedersen, M., Robins, V., Mortensen, K., and Kirkensgaard, J.J.K. (2020). Evolution of local motifs and topological proximity in self-assembled quasi-crystalline phases. Proceedings of the Royal Society A: Mathematical, Physical and Engineering Sciences *476*. https://doi.org/10.1098/rspa.2020.0170.
39. Jiang, F., Tsuji, T., and Shirai, T. (2018). Pore Geometry Characterization by Persistent Homology Theory. Water Resour. Res. *54*, 4150-4163. https://doi.org/10.1029/2017WR021864.
40. Dłotko, P., and Wanner, T. (2016). Topological microstructure analysis using persistence landscapes. Physica D: Nonlinear Phenomena *334*, 60-81. https://doi.org/10.1016/j.physd.2016.04.015.

41. Dixit, S., Crain, J., Poon, W.C.K., Finney, J.L., and Soper, A.K. (2002). Molecular segregation observed in a concentrated alcohol–water solution. Nature *416*, 829-832. https://doi.org/10.1038/416829a.
42. Soper, A.K., Dougan, L., Crain, J., and Finney, J.L. (2006). Excess Entropy in Alcohol−Water Solutions:  A Simple Clustering Explanation. The Journal of Physical Chemistry B *110*, 3472-3476. https://doi.org/10.1021/jp054556q.
43. Zaccarelli, E. (2007). Colloidal gels: equilibrium and non-equilibrium routes. J. Phys.: Condens. Matter *19*, 323101. https://doi.org/10.1088/0953-8984/19/32/323101.
44. Tsurusawa, H., Arai, S., and Tanaka, H. (2020). A unique route of colloidal phase separation yields stress-free gels. Science Advances *6*, eabb8107. https://doi.org/10.1126/sciadv.abb8107.
45. Stillinger, F.H. (1995). A Topographic View of Supercooled Liquids and Glass Formation. Science *267*, 1935-1939. https://doi.org/10.1126/science.267.5206.1935.
46. Suzuki, A., Miyazawa, M., Minto, J.M., Tsuji, T., Obayashi, I., Hiraoka, Y., and Ito, T. (2021). Flow estimation solely from image data through persistent homology analysis. Scientific Reports *11*, 17948. https://doi.org/10.1038/s41598-021-97222-6.
47. Thompson, E.P., and Ellis, B.R. (2023). Persistent Homology as a Heterogeneity Metric for Predicting Pore Size Change in Dissolving Carbonates. Water Resour. Res. *59*.
48. Savary, G., Guichard, E., Doublier, J.-L., and Cayot, N. (2006). Mixture of aroma compounds: Determination of partition coefficients in complex semi-solid matrices. Food Res. Int. *39*, 372-379. https://doi.org/10.1016/j.foodres.2005.09.002.
49. Déléris, I., Lauverjat, C., Tréléa, I.C., and Souchon, I. (2007). Diffusion of Aroma Compounds in Stirred Yogurts with Different Complex Viscosities. Journal of Agricultural and Food Chemistry *55*, 8681-8687. https://doi.org/10.1021/jf071149y.
50. Lee, Y., Barthel, S.D., Dłotko, P., Moosavi, S.M., Hess, K., and Smit, B. (2017). Quantifying similarity of pore-geometry in nanoporous materials. Nature Communications *8*, 15396. https://doi.org/10.1038/ncomms15396.
51. Krishnapriyan, A.S., Haranczyk, M., and Morozov, D. (2020). Topological Descriptors Help Predict Guest Adsorption in Nanoporous Materials. The Journal of Physical Chemistry C *124*, 9360-9368. https://doi.org/10.1021/acs.jpcc.0c01167.
52. Rosowski, K.A., Sai, T., Vidal-Henriquez, E., Zwicker, D., Style, R.W., and Dufresne, E.R. (2020). Elastic ripening and inhibition of liquid–liquid phase separation. Nature Physics *16*, 422-425. https://doi.org/10.1038/s41567-019-0767-2.
53. Curk, T., and Luijten, E. (2023). Phase separation and ripening in a viscoelastic gel. Proceedings of the National Academy of Sciences *120*, e2304655120. https://doi.org/10.1073/pnas.2304655120.

54. Relkin, P., Fabre, M., and Guichard, E. (2004). Effect of Fat Nature and Aroma Compound Hydrophobicity on Flavor Release from Complex Food Emulsions. Journal of Agricultural and Food Chemistry *52*, 6257-6263. https://doi.org/10.1021/jf049477a.
55. Saffarionpour, S. (2019). Nanoencapsulation of Hydrophobic Food Flavor Ingredients and Their Cyclodextrin Inclusion Complexes. Food and Bioprocess Technology *12*, 1157-1173. https://doi.org/10.1007/s11947-019-02285-z.
56. Debenedetti, P.G., and Stillinger, F.H. (2001). Supercooled liquids and the glass transition. Nature *410*, 259-267. https://doi.org/10.1038/35065704.
57. Wales, D.J. (2001). A Microscopic Basis for the Global Appearance of Energy Landscapes. Science *293*, 2067-2070. https://doi.org/10.1126/science.1062565.
58. Isayev, O., Fourches, D., Muratov, E.N., Oses, C., Rasch, K., Tropsha, A., and Curtarolo, S. (2015). Materials Cartography: Representing and Mining Materials Space Using Structural and Electronic Fingerprints. Chem. Mater. *27*, 735-743. https://doi.org/10.1021/cm503507h.
59. Ward, L., Agrawal, A., Choudhary, A., and Wolverton, C. (2016). A general-purpose machine learning framework for predicting properties of inorganic materials. npj Computational Materials *2*, 16028. https://doi.org/10.1038/npjcompumats.2016.28.
60. Jiang, X., Wan, D., Zheng, F., and Xie, Y. (2022). Ionization Equilibrium of Water Molecule Dominated Ethanol-water Binary Solution Self-assemble. Electrochemistry *90*, 067007-067007. https://doi.org/10.5796/electrochemistry.22-00054.